\documentclass[preprint,12pt]{elsarticle}
\makeatletter
\def\ps@pprintTitle{%
  \let\@oddhead\@empty
  \let\@evenhead\@empty
  \let\@oddfoot\@empty
  \let\@evenfoot\@empty}
\makeatother
\usepackage{graphicx}
\usepackage{amssymb}
\usepackage{braket}
\usepackage{mathtools}
\usepackage{float}
\usepackage{comment}
\usepackage{color}
\usepackage{subcaption}
\newcommand{\ketbra}[2]{\vert{#1}\rangle\langle{#2}\vert}
\newcommand{\cop}[1]{\hat{\sigma}_{#1}}

\begin{document}

\begin{frontmatter}

\title{Propagation Dynamics and Transient Amplification in Warm and Cold Atomic EIT Systems}

\author[uvic]{Andrew MacRae \corref{cor1}}
\ead{macrae@uvic.ca}
\author[carleton]{Connor Kupchak}

\cortext[cor1]{Corresponding author}

\address[uvic]{Department of Physics \& Astronomy, University of Victoria,
Victoria, BC V8W 2Y2, Canada}
\address[carleton]{Department of Electronics, Carleton University, Ottawa, ON K1S 5B6, Canada}

\begin{abstract}
We study the limitations on observing transient amplification in atomic systems exhibiting electromagnetically induced transparency (EIT) and evaluate the limits of optical Bloch equation (OBE) models. Using propagation-based Maxwell–Bloch simulations, we show that single-atom, spatially uniform OBE treatments overestimate gain by neglecting propagation dynamics. In two-level systems, this yields incorrect transmission, while in three-level systems, it predicts unrealistically large amplification. Furthermore, we show that Doppler averaging in warm vapor suppresses oscillatory ringing and the maximum achievable gain. Our results explain discrepancies between OBE predictions and experimental observations, and establish practical limits on transient gain in cold and thermally broadened EIT media.
\end{abstract}


\end{frontmatter}

\section{Introduction}
\label{sec:intro}
Electromagnetically induced transparency (EIT) is a well-understood effect in atomic, molecular, and optical (AMO) physics, first reported by Boller, Imamo\u{g}lu, and Harris~\cite{Harris1991} and applied in landmark slow light demonstrations~\cite{Hau1999, Liu2001}. Since then, it has been demonstrated in a variety of applications, ranging from optical storage for quantum memories~\cite{Eisaman2005} to precision magnetometry~\cite{Yudin2010} and classical signal processing~\cite{Li2024}.

EIT can be realized in different atomic energy-level configurations~\cite{Imamoglu1997,Kang2003,MacRae2008}. Typically, it involves a two-field interaction in which both fields are near-resonant with a common atomic transition. A strong \textit{control field} renders the atomic resonance transparent to a weak \textit{signal field}. In practice, this transparency is never perfect and is limited by atomic decoherence mechanisms, external field inhomogeneities, and relative phase instabilities between the involved fields. Moreover, the transparency window has a finite spectral width, which constrains the response time of the atomic system to propagating electromagnetic fields. A thorough understanding of the interplay between these dynamics and system parameters is essential for the success of EIT-based applications. For example, optical storage may require high efficiency~\cite{Ma2022}, while fast optical switching demands rapid response times~\cite{Bhushan2019}.

Several works have predicted~\cite{Harris1995} and reported~\cite{chen1998observation} transient gain during the interval required to establish steady-state EIT conditions. This atomic gain is important for two reasons. First, it provides the mechanism necessary for lasing without inversion (LWI), which has potential applications in nanophotonics~\cite{Werren2021}, nanolasers~\cite{Azzam2020}, and remote sensing~\cite{Richter2020}. Second, gain is often accompanied by spurious photons that can degrade optical fidelities and introduce excess noise. Thus, transient amplification in coherently prepared atomic ensembles has been a subject of sustained interest in the context of transient EIT.

Research on transient gain and LWI spans nearly three decades. In 1995, Zibrov \textit{et al.}~\cite{zibrov1995experimental} demonstrated LWI in rubidium vapor, observing a small but measurable steady-state gain. This was followed by semiclassical analyses~\cite{zhu1997light} and experimental demonstrations~\cite{chen1998observation} that revealed transient oscillations and predicted net transient gain under idealized conditions. These studies employed density matrix approaches to extract the optical response. Later works explicitly included propagation and Doppler effects~\cite{arve2004propagation}, presenting a unified picture of self-induced transparency, EIT and Raman transfer~\cite{clader2007two, wu2008evidence}. Experimental studies also reported transient oscillations in vapor cells enclosed within optical cavities~\cite{wu2008evidence}. Motivated by access to ultraviolet wavelengths, theoretical analyses have examined transient LWI in ladder and V-type systems~\cite{svidzinsky2013transient}. Subsequent experiments investigated EIT dynamics with Raman–Ramsey pulses~\cite{nikolic2015transient} and transient phase shifts~\cite{maynard2015time}. More recently, magnetically assisted transient gain was observed in a cold-atom cavity with a transverse magnetic field~\cite{carvalho2022lasing}.

In many of these studies, transient gain was inferred from an imaginary component of the susceptibility of less than zero. However, this approach neglects propagation effects that account for local phase and amplitude variations. In this work, we emphasize the importance of local propagation in determining the optical response of an atomic ensemble.

\section{Including Propagation in Atomic Simulations}
\label{sec:mbeexplain}

There are two standard semi-classical approaches to model the response of an atomic ensemble to an incident optical field: \emph{optical Bloch equations} (OBE) and  \emph{Maxwell-Bloch equations} (MBE) which we now outline.

\textbf{The OBE method} is a spatially uniform approach that calculates the response of a single atom and scales this solution based on the density and geometry of the atomic sample. Here, one starts with a Hamiltonian $\hat{H}$, describing the light-atom coupling, usually written in the rotating wave approximation and solves the master equation to obtain the time derivatives of the density matrix:

\begin{equation}
\label{eq:master_eq}
\frac{\partial \hat{\rho}}{\partial t}
= -\frac{i}{\hbar}\big[\hat{H},\hat{\rho}\big]
+ \sum_i \mathcal{L}_i[\hat{\rho}],
\end{equation}

\noindent where the terms $\mathcal{L}_i$ are the Lindblad operators that describe spontaneous emission and decoherence mechanisms. From the solutions to these equations and the atomic number density $N$, the polarization  $\mathbf{P} = N\bar{\mathbf{d}}$ is obtained from the expectation of the electric dipole operator $\hat{\mathbf{d}}$ for a particular transition: 

\begin{equation}\label{eq:meandipole}
\mathbf{P} = N \mathrm{Tr}(\hat{\rho} \hat{\mathbf{d}}) = N \mathbf{d}_{21}\rho_{12} + \text{H.c.}
\end{equation}

\noindent In Eq.~(\ref{eq:meandipole}) we take the ground (excited) state energy level to correspond to state $\ket{1}$ ($\ket{2}$). The terms $d_{21}$ and $\rho_{12}$ represent the optical dipole moment and density matrix element between these two states. Assuming a negligible nonlinear response, Eq.~\eqref{eq:meandipole} can be related to the susceptibility of a medium via  $\mathbf{P}=\varepsilon_0 \chi \mathbf{E}$, where here the susceptibility $\chi$ is a scalar quantifying the material response in the presence of an electric field. Defining the Rabi frequency as $\Omega \equiv \mathbf{d}_{21}\cdot\mathbf{E}/\hbar$, one obtains a scalar form of the linear susceptibility in terms of the density matrix solution to equation~\eqref{eq:master_eq}:

\begin{equation}
\chi = \frac{N |d_{21}|^2}{\varepsilon_0 \hbar}\,\frac{\rho_{12}}{\Omega}.
\label{eq:chi_from_rho}
\end{equation}

By taking the value of $\chi$ to be uniform throughout a medium of length $L$ and with a field of wavenumber $k=2\pi/\lambda$, the field propagation can be modeled using the Beer--Lambert law. Using $n = \sqrt{1+\chi}\approx1+\frac{\chi}{2} = (1+\frac{1}{2}\mathrm{Re}\left[\chi\right]) + \frac{i}{2}\mathrm{Im}\left[\chi\right]$, this is written as:

\begin{equation}
\Omega(z) = \Omega(0) e^{i\left(1+\frac{\mathrm{Re}\left[\chi\right]}{2}\right)kz} e^{-\frac{\mathrm{Im}\left[\chi\right]}{2}kz},
\label{eq:beer}
\end{equation}

\noindent where $\mathrm{Re}\left[\chi\right]$ and $\mathrm{Im}\left[\chi\right]$ correspond to the real and imaginary components of the susceptibility respectively.

\textbf{The MBE approach} provides a mathematical formalism that naturally integrates propagation through the medium and introduces a coupled set of solutions for the atoms and optical fields. Here, the optical field is treated under the paraxial and slowly varying envelope approximation such that

\begin{equation}
\frac{\partial \Omega(z,t)}{\partial z} + \frac{1}{c}\frac{\partial \Omega(z,t)}{\partial t}
= i\eta\rho_{12}(z,t),
\qquad
\eta \equiv \frac{\omega N |d_{21}|^2}{2\varepsilon_0 \hbar c}.
\label{eq:mbe}
\end{equation}

\noindent By translating to a co-moving frame $(\zeta=z,\;\tau=t-z/c)$, Eq.~\eqref{eq:mbe} reduces to:

\begin{equation}
\frac{\partial \Omega(\zeta,\tau)}{\partial \zeta} = i\eta\,\rho_{12}(\zeta,\tau), \label{eq:prop}
\end{equation}

In the weak-probe, steady-state limit, $\rho_{12}\propto \Omega$ and neither $\rho_{12}$ nor $\Omega$ vary appreciably with $z$, so the MBE reduces to the OBE result in Eq.~(\ref{eq:beer}). Beyond the weak-probe regime, transient propagation becomes important, especially at high optical depth. In this case, a full MBE treatment is required.

\section{Importance of Including Propagation: Two-Level Atoms}
\label{sec:2LAex}
\begin{figure}
    \centering
    \includegraphics[width=\linewidth]{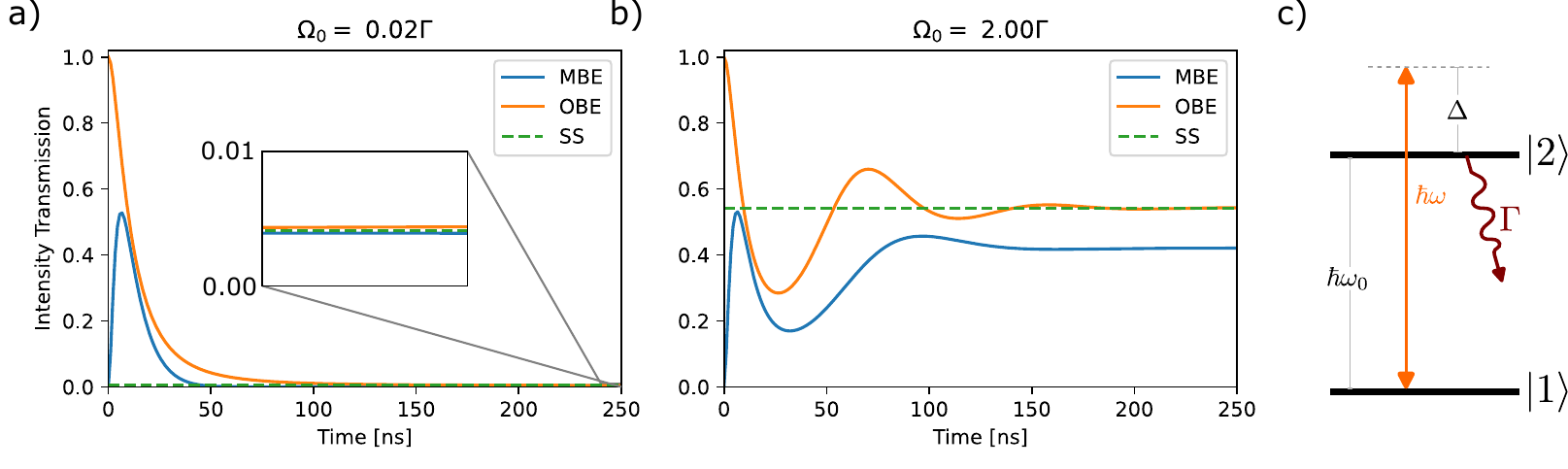}
    \caption{Transmission of a signal through a two-level atom for the case of a weak field ($\Omega = 0.02\Gamma$, a) and a strong field ($\Omega = 2\Gamma$, b). For the case of the strong field, the diminished saturated absorption during propagation (MBE, blue) results in less transmission than predicted by the spatially uniform approximation (OBE, orange). Panel (a) shows a zoom-in for longer times. The green dotted line shows the steady-state solution to the system by setting the time derivatives in Eq.~\eqref{eq:master_eq} to zero. Here, the optical depth is $5.5$ and the medium is a $1$~mm cloud of non-Doppler broadened Rb atoms. (c) The atomic model used in this simulation.}
    \label{fig:2LA_MBEvsOBEvsSS}
\end{figure}

We now note the effect of propagation in a two-level atomic model, sketched in figure~\ref{fig:2LA_MBEvsOBEvsSS}c. The Hamiltonian for such an atom can be written in the rotating wave approximation as $\hat{H} = -\hbar\left(\Delta\cop{22} + \frac{\Omega}{2}\cop{12}+ \frac{\Omega^*}{2}\cop{21}\right)$. Here $\cop{ij}\equiv\ketbra{i}{j}$ are atomic transition operators written in the interaction picture.

In an optically driven two-level atom, one obtains closed-form solutions to the steady-state OBEs, which via Eq.~\eqref{eq:chi_from_rho} yields a susceptibility given by~\cite{siddons2014light}

\begin{equation}
\chi(\Omega,\Delta)
= -\frac{N \vert d_{21}\vert^2}{\varepsilon_0 \hbar} 
\frac{2\Delta - i\Gamma}{4\Delta^2 + \Gamma^2 + 2\Omega^2}.
\label{eq:chi_2la}
\end{equation}

\noindent Here, $\Delta=\omega-\omega_{21}$ and $\Gamma$ is the spontaneous emission rate from $\ket{2}$ to $\ket{1}$. 

In the limit of a weak field, $4\Delta^2+\Gamma^2 \gg 2\Omega^2$, the $\Omega^2$ term is negligible and we can simplify Eq.~(\ref{eq:chi_2la}) to 

\begin{equation}
\chi(\Omega,\Delta)
= -\frac{N |d_{21}|^2}{\varepsilon_0 \hbar} 
\frac{1}{2\Delta + i\Gamma}.
\label{eq:chi_2laweak}
\end{equation}

\noindent Note that on resonance, the susceptibility is purely imaginary and thus Eq.~\eqref{eq:beer} yields:
\begin{equation}
    \label{eq:resoptdepth}
    \vert\Omega(z)\vert^2 = \vert\Omega_0\vert^2e^{-\text{Im}[\chi]kL} =  \vert\Omega_0\vert^2e^{-\text{OD}},
\end{equation}

\noindent where OD $= \frac{N \vert d_{21}\vert^2}{\varepsilon_0 \hbar\Gamma}kL$ is the resonant optical depth. In the regime of low optical depth, OD$\ll1$, the field $\Omega$ will not be considerably altered during propagation and the spatially uniform approximation of the OBEs will be valid. At high optical depths, when saturation is present, the field undergoes depletion during propagation and the nonlinear saturation effects due to the $\Omega^2$ term in Eq.~\eqref{eq:chi_2la} will lead to a spatially dependent absorption coefficient.

This effect is illustrated in Fig.~\ref{fig:2LA_MBEvsOBEvsSS} where we assume a $1~$mm diameter spherical magneto-optical trap of atomic density $5\times10^{16}~\mathrm{atoms}/\mathrm{m}^3$, and on-resonance optical depth $2.0$. Fig.~\ref{fig:2LA_MBEvsOBEvsSS}(a) shows the results of a low Rabi frequency ($\Omega = \Gamma/50$) and negligible saturation that corresponds to similar steady-state transmissions between the OBE and MBE methods. In contrast, figure~\ref{fig:2LA_MBEvsOBEvsSS}(b) shows a situation where the susceptibility becomes saturated ($\Omega = 2\Gamma$). In this scenario, the OBE treatment assumes that all absorption takes place at a single location and predicts higher transmission than the MBE result. The more realistic MBE simulation shows that 
as the field propagates, its local intensity decreases, reducing saturation and thereby increasing the local absorption later in the medium. This feedback effect leads to significantly less transmission than would be expected from the naive OBE and steady-state approaches.

On the other hand, in a situation where the weak-signal response exhibits gain (i.e. $\mathrm{Im}~\chi<0$), the amplitude of the optical field will increase during propagation and in turn significantly modify the index of refraction. In this scenario, the OBE treatment will overestimate the net amplification, while the MBE will predict an output more closely matched to experimental observation. This is discussed in the next section.

\section{Limitations on Transient Gain in EIT Systems}
\label{sec:MBEinEIT}

\begin{figure}
    \centering
    \includegraphics[width=\linewidth]{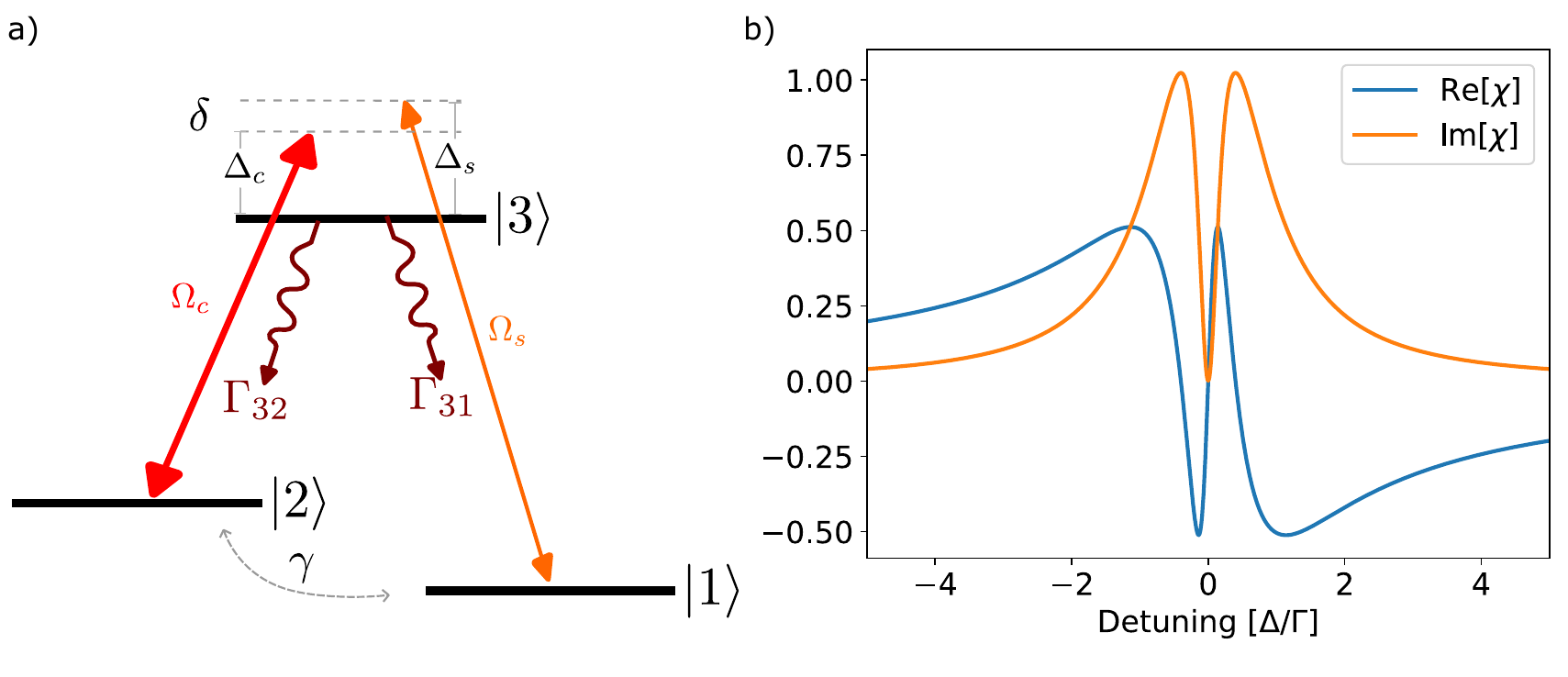}
    \caption{(a) $\Lambda$-type  EIT configuration where $\ket{1}$ and $\ket{2}$ are metastable states with no dipole coupling between them, sharing a common excited state $\ket{3}$. Population decay occurs from state $\ket{3}$ to states $\ket{1}$ and $\ket{2}$ at rates $\Gamma_{31}$ and $\Gamma_{32}$ respectively. Decoherence mechanisms between states $\ket{1}$ and $\ket{2}$ are modeled as a ground state dephasing term $\gamma$. (b) Calculated real and imaginary parts of the signal field susceptibility.}
    \label{fig:EITscheme}
\end{figure}

The spatially uniform OBE approach can be particularly unreliable in light-atom regimes where gain is present. To exemplify this, we study EIT in an atomic $\Lambda$-system in the presence of a strong control field $\Omega_c$ and weak signal field $\Omega_s$. The Hamiltonian for the system can be written as

\begin{equation}
    \hat{H} = -\hbar\left(\delta\cop{22} + \Delta_s\cop{33} + \frac{\Omega_s}{2}\cop{13} + \frac{\Omega_c}{2}\cop{23} + \frac{\Omega_s^*}{2}\cop{31} + \frac{\Omega_c^*}{2}\cop{32}\right)
    \label{eq:EIThamiltonian}
\end{equation}

\noindent where $\Delta_s$, $\Delta_c$ are the signal and coupling field one-photon detunings respectively, and $\delta\equiv\Delta_s-\Delta_c$ is the two-photon detuning.

The susceptibility can be predicted by solving Eq.~\eqref{eq:EIThamiltonian} using the same approach described in Section~\ref{sec:2LAex}. Regions of gain are predicted when the transient response shows $\mathrm{Im}\left[\chi\right]<0$ and for large optical depths, Eq.~\eqref{eq:beer} correlates with a larger transient gain. 

When propagation effects are considered in the MBE treatment, the amplitude of the signal field can increase and accumulate phase due to the dispersion effects from the medium. In this case, the underlying assumption of $\rho_{12}\propto\Omega_s$ is no longer valid and nonlinear components of the susceptibility such as non-negligible $\chi^{(3)}$ values and population redistribution become apparent. This interplay reduces the overall magnitude where  $\mathrm{Im}~\chi<0$ and the control field can no longer be assumed constant. These atomic-optical feedback effects must be modeled and tracked in the propagation calculations.

\begin{figure}
    \centering
    \includegraphics[width=\linewidth]{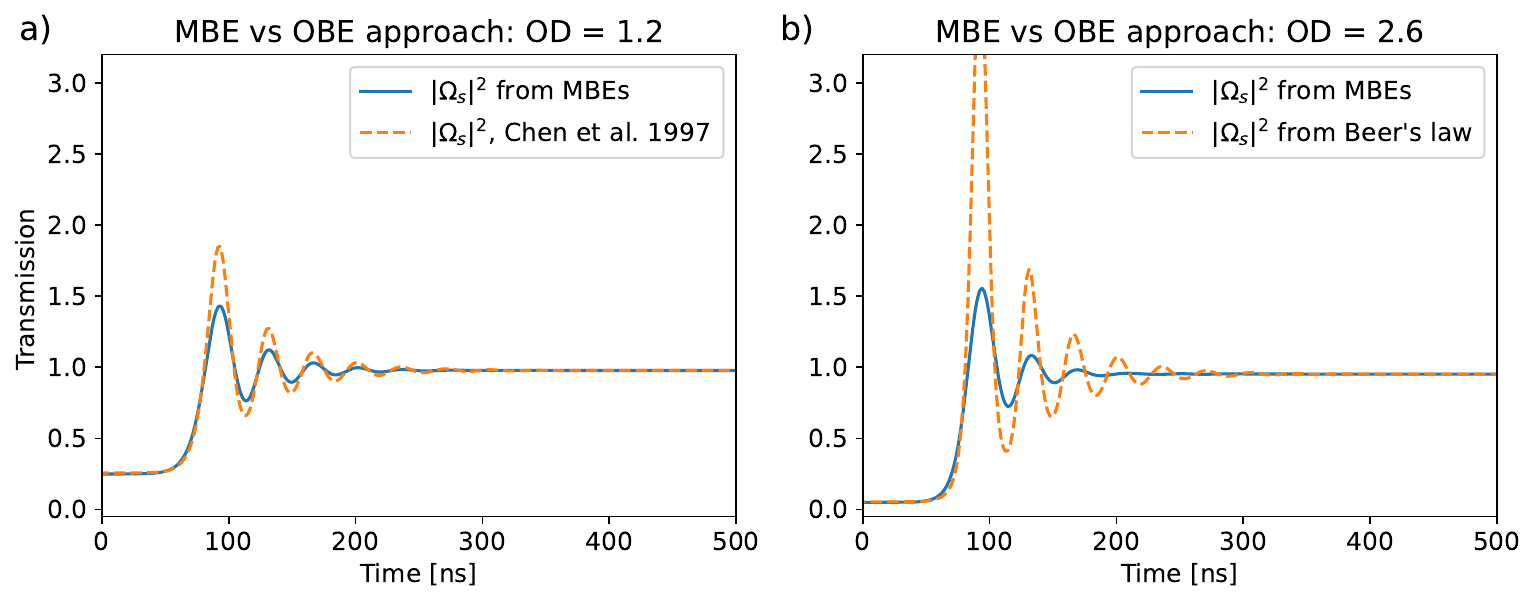}
    \caption{Comparison between MBE and OBE simulations. (a) Parameters from Ref.~\cite{chen1998observation}: $\Gamma=2\pi\times5.86~\mathrm{MHz}$, $\Omega_c=1.6\Gamma$, $\Omega_s=0.1\Gamma$, $\gamma=0.3\Gamma$, $N=2\times10^6$ atoms in a $1.2~\mathrm{mm}$ sphere (OD $\approx 1.2$). (b) Same parameters but with an increase in OD to $\approx 2.6$, illustrating the growing discrepancy. Peak gains are $3.56$ (OBE) vs $1.59$ (MBE).}
    \label{fig:MBEvOBE_highOD_combined}
\end{figure}

The significance of this discrepancy is highlighted by the experimental investigation of transient EIT reported in~\cite{chen1998observation}. The authors stated that while gain was not reported for their maximum observed coupling field strength ($\Omega = 1.6\Gamma$), it would be present for increased Rabi frequencies near $\Omega = 10\Gamma$. Figure~\ref{fig:MBEvOBE_highOD_combined} compares our MBE treatment to the OBE treatment using the parameters of~\cite{chen1998observation}: $\Gamma=2\pi\times5.86~\mathrm{MHz}$, $\Omega_c = 1.6\Gamma$, $\Omega_s = 0.1\Gamma$, $\gamma = 0.3\Gamma$. Fig.~\ref{fig:MBEvOBE_highOD_combined}(a) shows a direct match to the report, with $2\times10^{6}$ atoms in a $1.2$~mm spherical cloud of cold Rb atoms. Fig.~\ref{fig:MBEvOBE_highOD_combined}(b) shows a modest increase to $12\times10^6$ atoms in the same cloud, a number easily achieved with modern magneto-optical traps. Our results show that optical propagation based MBE treatment still predicts gain comparable to~\cite{chen1998observation}, albeit reduced. However, when the optical depth is increased to values within experimental limits, the peak transient gain becomes very large. For example, in a compressed MOT with $\mathrm{OD}=8$, the OBE approach predicts gains of $G = \Omega_\mathrm{out}/\Omega_\mathrm{in} > 800$, whereas the transient MBE gain remains around $1.25$.

This discrepancy between methods can be understood by the discussion in Section~\ref{sec:2LAex}. During signal field propagation, the amplitude grows sharply and experiences a phase shift. The atoms at a particular $z$ position thus interface with a field exhibiting time-dependent amplitude modulation and phase chirp. This modifies the system such that the ideal EIT conditions of weak signal field $\Omega_s\ll\Omega_c$ and narrow pulse bandwidth $\Delta t_s \gg \Delta\nu^{-1}$ are not valid. For large gains, the proportionality between the susceptibility and $\Omega_s$ no longer holds but coherent population trapping remains possible. In Fig.~\ref{fig:MBEvOBE_highOD_combined}(a) where OD$\approx1$, the nonlinearities are negligible but have a more apparent impact on field propagation at higher optical depths.

\section{Transient Gain in Doppler-Broadened Systems}
\label{sec:DoppEIT}

We now examine limitations of the OBE approaches in a Doppler broadened system. Here, atoms of mass $m$ are in a thermal bath at temperature $T$ with velocities described by Maxwell--Boltzmann distribution.

\begin{figure}
    \centering
    \includegraphics[width=\linewidth]{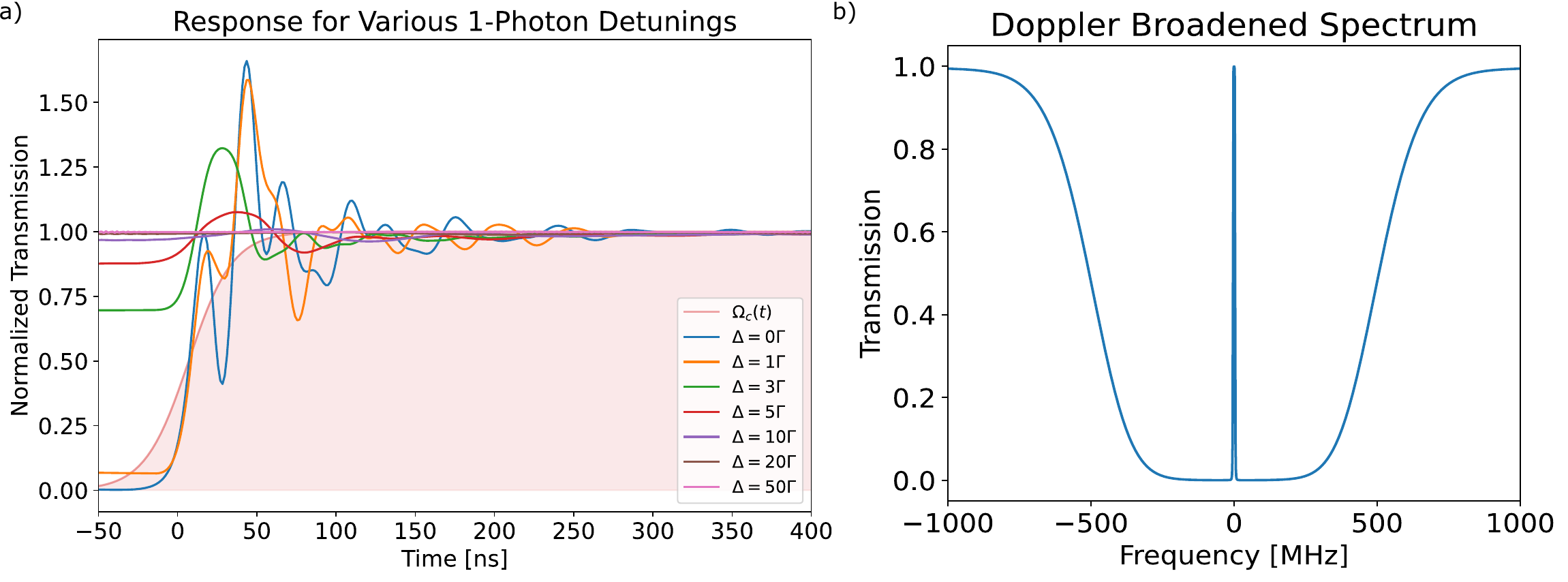}
    \caption{(a) Transient response for different velocity subclasses in factors of the spontaneous emission $\Gamma$. (b) Steady-state, Doppler-broadened EIT profile obtained by integrating over the full velocity distribution.}
    \label{fig:MBEvOBE_multiDelta}
\end{figure}

\begin{equation}
    P(v)dv = \frac{1}{\sqrt{2\pi}\sigma_v}e^{-\frac{1}{2}\left(\frac{v}{\sigma_v}\right)^2}dv, \qquad \sigma_v^2 = \frac{k_BT}{m},
    \label{eq:dopp_dist}
\end{equation}

\noindent with $k_B$ being the Boltzmann constant. Atoms will experience a Doppler shift based on their velocity $v$ of $\Delta_d = kv$, with $k\equiv2\pi/\lambda$ being the vacuum wavenumber. The corresponding one-photon detuning terms for the signal and control fields then become:

\begin{equation}
\Delta_s(v)=\Delta_s - k_s v, \qquad\Delta_c(v)=\Delta_c - k_c v.
\end{equation}

For two-level atoms, this results in an overall broadening of the absorption profile, but for EIT in a collinear propagation regime, the effect is more subtle. Since the signal and coupling field wavenumbers are similar $|k_s-k_c| \ll k$, the two-photon resonance becomes Doppler-narrowed~\cite{finkelstein2023practical}. In the dressed state picture, each detuning corresponds to an effective Rabi frequency, given by $\Omega_{\mathrm{eff}}(\Delta) \approx \sqrt{\Omega_c^2 + 4\Delta^2}$. This leads to the simultaneous presence of different oscillation frequencies due to the collection of velocity subclasses. This collection of transient behaviors is illustrated in Fig.~\ref{fig:MBEvOBE_multiDelta}(a).

\begin{figure}
    \centering
    \includegraphics[width=0.7\linewidth]{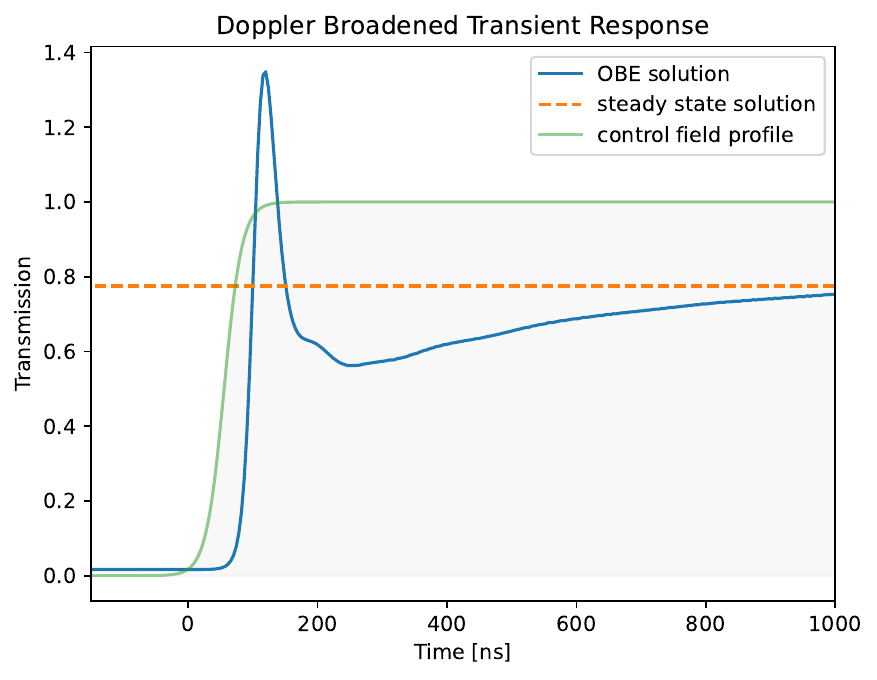}
    \caption{Transient response calculated via the OBEs in a Doppler broadened medium. Simulation parameters consist of a $2.5~$cm vapor cell at $ T = 60^{\circ}\mathrm{C}$ with $\Delta_s = \Delta_c = 0$, $\Omega_c = 2\Gamma$ and $\gamma = 0.001\Gamma$.}
    \label{fig:OBE_incldoppler}
\end{figure}

The time dependent response can be calculated by integrating the Doppler velocity probability distribution Eq.~\eqref{eq:dopp_dist} along the response for a given velocity subclass~\cite{pizzey2022laser} such that:

\begin{equation}
    \rho_D(\Delta_s,\Delta_c,t) = \int_{-\infty}^\infty P(v) \rho(\Delta_s-kv,\Delta_c-kv,t)dv.
    \label{eq:doppler_soln}
\end{equation}

\noindent Eq.~\ref{eq:doppler_soln} is applied to the MBEs of Eq.~\eqref{eq:mbe} or directly converted to a susceptibility via Eq.~\eqref{eq:chi_from_rho}. Fig.~\ref{fig:MBEvOBE_multiDelta}(b) shows the resulting steady state Doppler broadened EIT profile expected for a $2.5$~cm Rb vapor cell at T = $60^\circ$~C and $\Omega_c = 15~\Gamma$.

The immediate consequence of Doppler broadening is that the continuum of different transient oscillation frequencies effectively combine to wash out individual oscillations characteristic of cold atom systems.

Figure~\ref{fig:OBE_incldoppler} shows the effect of this integration over the Maxwell-Boltzmann distribution and solving the OBEs for the system. Here, we again simulate a $2.5~$cm vapor cell at  T = $60^{\circ}\mathrm{C}$. It can be seen that the oscillations of the signal field are suppressed in addition to a large reduction in total amplification that would be present in a cold atomic system. We can conclude that the presence of thermal Doppler broadening in vapor cell systems severely constrains the observation of transient gain that is more readily observed in a cold atom system.

\section{Conclusions}
\label{sec:conclusions}
In conclusion, we have presented two main causes that restrict the observation of high-gain, transient responses in atomic systems: field propagation effects and Doppler broadening. Single-atom, spatially uniform OBE models are adequate in the weak-probe, steady-state regime but become unreliable when the impact of the atoms on the propagating field needs to be considered.  This becomes immediately apparent in moderate-to-high optical depth media and when there are time-dependent responses from the atoms. In two-level media, an assumption of the medium exhibiting uniform saturation will result in an overprediction of the signal field transmission. The spatially uniform OBE neglects that the probe intensity (and therefore saturation) decreases along $z$, which restores absorption downstream. Accounting for this yields lower transmission than OBE predicts.

In a three-level $\Lambda$ system, inferring the gain solely from a negative local susceptibility can also lead to inaccuracies. As the gain in the signal field increases, nonlinearities, control field depletion and transient phase shifts will reduce this amplification. In a Doppler broadened system, the presence of various velocity subclasses will result in averaging the atomic coherence that effectively washes out the probe ringing and amplification expected in a non-Doppler system. In typical atomic vapor-cell conditions (e.g., $60^{\circ}$C, 25~mm), the combined effects of Doppler averaging and propagation feedback lead to a large suppression in the transient gain that is predicted in non-propagating OBE models.

Practically, our results recommend: (i) using Maxwell-Bloch simulations when evaluating systems characteristic of transient amplification or high optical depths, (ii) treating negative $\mathrm{Im}\,\chi$ as a necessary but not sufficient indicator of the presence of net gain and (iii) mitigating the impact of velocity distributions or enhancing mode selectivity (cold atoms, optical cavities) if a sizable transient amplification is required. Future work will probe extensions of these models to multi-level hyperfine structures, transverse effects and cavity-enhanced geometries in both cold and thermal ensembles.

\section{Acknowledgments}
We acknowledge support of the Natural Sciences and Engineering Research Council of Canada (NSERC) (grant RGPIN-2021-03289).

\bibliographystyle{elsarticle-num}
\bibliography{propEITgain}

\end{document}